\documentclass[english,11pt]{article}

\usepackage{amssymb}
\usepackage{graphicx}
\usepackage{amsfonts}
\usepackage{bbm}
\usepackage[T1]{fontenc}
\usepackage[latin1]{inputenc}
\usepackage{babel}
\usepackage{color}

\newcommand{\be}{\begin{equation}}
\newcommand{\ee}{\end{equation}}
\newcommand{\rf}[1]{(\ref{eq:#1})}

\begin{document}

\title{Quaternion-Loop  Quantum Gravity}

\author{ M. D. Maia\footnote{maia@unb.br}, \hspace{2mm}  S. S e Almeida
Silva\footnote{su8274@gmail.com}  \hspace{1mm}\&\hspace{1mm} F. S.
Carvalho\footnote{fabiosalgado@gmail.com}  \\ Universidade de Bras\'{\i}lia,
Instituto de F\'{\i}sica, Bras\'{\i}lia, DF 70919-970}

\maketitle

\begin{abstract}
 It is shown   that    the  Riemannian  curvature  of  the 3-dimensional hypersurfaces in space-time,  described  by the Wilson loop integral, can be  represented by a  quaternion quantum operator induced by  the  SU(2) gauge  potential, thus providing  a  justification for quaternion quantum gravity  at the Tev  energy  scale.
\end{abstract}

\tableofcontents

\section{Introduction}

In one of his  earliest   attempts  towards  a  canonical  formulation of  the  gravitational field,   Dirac described the propagation  of   3-dimensional space-like hypersurfaces in  space-time along an orthogonal time-like  direction.  The resulting non-zero Hamiltonian allowed the  construction of  the canonical   equations,   whose solutions describe the  evolution of  a special  space-time  foliation given by  by 3-dimensional space-like  hypersurfaces \cite{Dirac1}. Since  that procedure  is  not   compatible  with   the diffeomorphism  invariance  of    general  relativity,  the next logical step  was to define  a   covariant foliation. For  that purpose Arnowitt, Deser  and  Misner considered    an  arbitrary   propagation  direction  in space-time, with  a time-like   lapse component $N_0$,  and   three   space-line components  $N_i$ composing a shift vector \cite{ADM}.  Unfortunately,    the presence of the shift vector   implies   that  the Hamiltonian  vanishes,   thus frustrating  the  intended covariant canonical  formulation.  The  problem persists
even after applying   Dirac's   procedure  for  constrained systems \cite{Dirac2},    essentially  because      the   Poisson bracket  structure  is not invariant under the  group of  diffeomorphisms  \cite{Kuchar}.

In view  of  such negative  results,  it seemed for  a  while   that   a  covariant  canonical   quantum gravity  program   describing  the quantum  fluctuations  of  3-hypersurfaces  in  space-time  was not possible.  However  an  interesting  alternative was  proposed by  A.  Ashtekar,  suggesting  that a  quantized  $SU(2)$  gauge  field acting as  an  auxiliary  variable,  may  induce  the  quantization of the gravitational field  \cite{Ashtekaro}.  Indeed,   the Riemann curvature  of the 3-dimensional    hypersurfaces can be  written  in terms of the affine connection  of   the group  of holonomy of  triads along a closed loop,  as given   by  the Wilson loop integral \cite{Ashtekar,Smolin}.  Since   the holonomy group  is  isomorphic  to  $SO(3)$,  which  in turn is     isomorphic  to the  gauge group   $SU(2)$,  then the quantized  $SU(2)$ gauge  field  induces the quantization of the  3-curvature.
However, the  mathematical  difficulties    associated  with the  expansion of   the Wilson  integral  including the spin operators after the  quantization   (ie the Mandelstam inequalities),  eventually led to  a  different  route:  Since  the  $SU(2)$  operators   act  on two-component   spinor  fields,  then    the curvature of  the  3-dimensional hypersurfaces can  also be  expressed     in terms  of   an  underlying projective  space  defined  at the Planck  scale, as  a discrete    structure such as Penrose's   spin network.  The present stage of  the development of  loop  quantum gravity,    based  on such  spin network and  using  finite difference  equations,  is   a current  mainstream  on  quantum gravity  \cite{Smolin,Han}.

The purpose of this  paper is  to present   an   alternative    to loop quantum gravity, based on the fact that  the  expression of the 3-dimensional  Riemann  curvature in terms of the Wilson  integral   is  independent  of  the spinor  discrete  structure. We   use the  fact that the   quaternion algebra is  more  fundamental than any  of  its representations,  including  the spinor   representations.  This last property allows us  to  make  an   explicit  association between   the  curvature given by the  Wilson  loop integral and  quaternion  quantum mechanics,  naturally  suggesting   the  name of quaternion-loop quantum gravity. The main difference with  the standard  loop  quantum gravity being that  instead  of using a   discrete  projective  space structure, quaternion-loop quantum gravity  is  based on the standard  continuum   analysis
of  manifolds.
In the  next  section   we  give  a  very brief  review of  properties of  quaternion  and its  2-component spinor  representation. In section 3  we  describe the  curvature of a   3-dimensional hypersurface  in terms of  the  Wilson  loop  integral and  use the   quantization of the   $SU(2)$  gauge  field  to  express  that   curvature   as  a  quaternion quantum  operator.  At the end  we  give  the example  of the  3-dimensional  curvatures  induced by instanton and anti-instanton  gauge  fields.

\section{Quaternions and Spinors}

 A common textbook  explanation on the necessity of the complex formulation of  quantum  mechanics is that it is  required  to   guarantee the completeness  of  the spectrum of eigenvalues of the  Casimir spin operator \cite{Chevalley}. However, since  2-component spinors  are  vectors belonging to  the  representation space of
a representation of the group of  automorphisms  of the  quaternion algebra,  the above  mentioned  justification for   complex   quantum mechanics    is  really  a justification for  quaternion   quantum mechanics.
It has been  suggested  that the    non-commutativity  of the quaternion  algebra may   detail  some  properties of    quantum  fields  and  quantum geometry,  which  are  not manifestly present  in the simpler  complex quantum theory \cite{Finkelstein, Adler,Adler1}.

The  quaternion   algebra   is  a   Clifford  algebra  with two  generators  $\{e_1,e_2 \}$  satisfying a  multiplication table $e_{i}e_{j} + e_{j}e_{i} = 2g_{ij}e_{0},\;\;\;
e_{0}e_{i} =e_{i}e_{0},\;\;\;
e_{0}e_{0}=e_{0}$,  where  $g_{ij}$ are the coefficients of a  given quadratic form.  It  follows that the  number of independent  products  of the  generators    is  four,    defining  the   dimension of the algebra. Denoting   $e_3 =e_1 e_2$,  a general  quaternion  can  be   written  as
\[
X= X^{0}e_{0} +X^1 e_{1} + X^{2}e_{2}   + X^3 e_3
\]
The multiplication table is  invariant under the group of automorphisms   $e'_\alpha =u  e_\alpha  u^{-1}$,  which is      isomorphic   to   the  $SO(3)$  group.  The  conjugate of  a quaternion is  defined by
$
 \bar{e}_{i} =-e_{i},\;\; \bar{e}_{0}=e_{0}
$,
so that  the   real part of  a quaternion is  $Re(X)  =(X + \bar{X})/2$, and  the  imaginary (or  vector) part  is   $Im (X) = (X -\bar{X})/2$.  Therefore,  quaternions  have  a norm  $||X||^{2}=X\bar{X}$,  and
for   $g_{ij} =\delta_{ij}$  the norm   has the  Euclidean  expression
\[
||X||^{2}=X_{0}^{2} +X_{1}^{2}+X_{2}^{2} +X_{3}^{2}
\]
The  existence of  such norm  allow  us  to  construct   the   inverse of  a quaternion as $X^{-1}  = \bar{X}/||X||^2$.  It  also  allows  to   write  $||XY||  =||X|| ||Y||$,  a property of the quaternion algebra which is  shared  with  the  real  and   complex algebras.  In fact  these are  the only known associative normed  division algebras,  a   necessary  condition  for the  definition of the   standard  mathematical  analysis, including the  notions of limit,  continuity and  derivative  of  a  quaternion function $F(X)$ of  a  quaternion variable $X$. While in general  we have that the product of  variations $\Delta F \Delta X^{-1} \neq  \Delta X^{-1} \Delta F$, the  division property  says that
\be
\lim_{\Delta X \rightarrow 0}\left|\left| \Delta  F (X)  (\Delta X)^{-1}\right|\right| =
\lim_{\Delta X \rightarrow 0}\left|\left| (\Delta X)^{-1} \Delta  F (X) \right|\right|=
 \lim_{\Delta X
\rightarrow 0} \frac{||\Delta  F  (X)|| }{||\Delta X ||}  \label{eq:x}
\ee
so that although  the  limit operations are  well defined,  the   left and  right  derivatives need not be  equal.
If  we  impose that they are,  then the  analyticity   conditions
become  too  restrictive  to  allow for  the definition of
  some relevant  analytic  functions \cite{Maia}.

\vspace{4mm}
The  spinor representations  of the quaternion  algebra
starts  with   the  Pauli matrices
\begin{equation}
\begin{array}{l} \sigma_{0}=\left(
            \begin{array}{ll}
            1 \;\;\;\; 0\\
            0 \;\;\;\; 1
            \end{array}
            \right)
\;\;\; \sigma_{1}=\left(
            \begin{array}{ll}
            0 \;\;\;\;1\\
            1 \;\;\;\; 0
            \end{array}
            \right)
\;\;\; \sigma_{2}= \left(
          \begin{array}{ll}
          0 \; -i\\
         i \;\;\;\;\; 0.
          \end{array}
          \right)
\;\;\, \sigma_{3}=\left(
          \begin{array}{ll}
          1 \;\;\;\;\; 0\\
          0\; -1
          \end{array}
          \right)
\end{array}  \label{eq:Pauli}
\end{equation}
which satisfy (for  $g_{ij}=\delta_{ij}$)  the  same quaternion multiplication table
$
\sigma_{i}\sigma_{j} + \sigma_{j}\sigma_{i} = 2\delta_{ij}\sigma_{0},\;\;\;
\sigma_{0}\sigma_{i} =\sigma_{i}\sigma_{0},\;\;\;
\sigma_{0}\sigma_{0}=\sigma_{0}  \label{eq:table}
$.
In this representation  a general  quaternion corresponds to  a  2x2 matrix
\be
 X=X^{0}\sigma_{0} +
X^{1}\sigma_{1} + X^{2}\sigma_{2} + X^{3}\sigma_{3}=
\left(
\begin{array}{cc}
  X^0 +X^3& X^1 - iX^2\\
X^1 + iX^2 & X^0 -X^3 \\
\end{array}
\right)  \label{eq:M}
\ee
Reciprocally,  any 2x2 complex  matrix  operator acting on the 2-component spinor  space  can  be also written as  a   quaternion operator.
A quaternion  $X$   can also  be  expressed in terms of  two-component spinors as  $X  =  X^\mu \sigma_\mu =  X^\mu \sigma_{\mu A}{}^{\bar{B}} \, \Psi^A \Psi_{\bar{B}}$
where the spin-tensor $\sigma_{\mu A}{}^{\bar{B}}$  are  the components of  each of  the  matrices $\sigma_\mu$,    $\Psi_A$  are the   components of the
2-component spinors  and  $\Psi^{\bar{A}}$ denotes  its  Hermitian conjugate. Therefore,   a  quaternion  can  be seen as determined by  a  more  elementary   spinor structure.
Using a  combinatorial  analysis associated  with the  eigenvalues of the spin Casimir operator, Penrose postulated   a discrete   projective space,   the spin  network, which is  a mathematical
formulation of  such  underlying structure \cite{Penrose}.

\section{Quaternion-Loop   Quantum Gravity}
The  curvature   of the   3-dimensional hypersurface  in the Riemann  sense  can  be  evaluated by  transporting  a tangent vector  field  along a closed  loop consisting of
a  single  continuous curve   $\gamma$,  starting at  a point  $p$  and  ending in  the same  point.  Comparing the result  with the  original  vector field at  $p$,  their difference can be expressed by the Wilson loop integral \cite{Adler1,Modanese}
\be
R(\Gamma)  =  P  e^{\oint_\gamma \Gamma_i dx^i }, \;\; i=1..3
\label{eq:loop}
\ee
where   the  coefficients  $\Gamma_i$  are   the  components of the   affine  connection $\nabla$  evaluated in a reference frame
(a triad) defined in the    hypersurface. The motion of  the triad  along the  loop  characterizes the  triad  holonomy group  which is  isomorphic  to $SO(3)$. The  ordering   factor $P$  can  be  thought  of  as  an  integration constant.
Admitting that  $\Gamma_i$  are   continuous   functions of the coordinates  along the loop, the     exponential  in  \rf{loop} is  a well defined   real analytic  function  represented  by the standard exponential converging positive power  series
\be
  e^{\oint_\gamma \Gamma_i dx^i }=
 1  + \oint_\gamma \Gamma_i dx^i   +(\oint_\gamma \Gamma_i dx^i)^2 +\cdots  \label{eq:realexpntl}
\ee

The fact that  the  group of  automorphisms  of the  quaternion algebra  is  also  isomorphic  to  $SO(3)$, which is in turn  isomorphic  to $SU(2)$,  suggests  a classical-quantum
correspondence   between  the curvature  of the  3-dimensional  hypersurfaces in  space-time,   expressed   as a  quaternion operator and  the $SU(2)$ gauge  field.
Since $\Gamma_i$ are the components of  the connection of  the   triad  holonomy group,   we may  establish  this  correspondence as \be
\Gamma_i  \longleftrightarrow  i\hbar  \, \hat{\mathcal{A}}_i, \;\; i=1..3
\label{eq:quantization}
\ee
where  $\hat{\mathcal{A}}_i$  are  the  3-dimensional  components of the quantized $SU(2)$  gauge potential.
Replacing this  in   \rf{loop},  we obtain the  quantum curvature operator of the  3-dimensional  hypersurface
\be
\hat{\mathcal{R}}({\cal A}) =   \mathcal{P}e^{i\hbar \oint_\gamma \hat{\mathcal{A}}_i dx^i }, \;\; i=1..3  \label{eq:R}
\ee

However, as   it was  already commented, the analyticity of a quaternionic  exponential function as   in \rf{R} is not well  defined.   Still,   it is possible  to   apply \rf{quantization}
to each  term  of the  classical  expansion in the right hand side of  \rf{realexpntl},  thus obtaining  a  series of  quaternion polynomials,  which  converge to the  quaternion  exponential \cite{Maia}:
\be
e^{i\hbar \oint_\gamma \hat{\mathcal{A}}_i dx^i }\stackrel{def}{=}\sigma_0  + i\hbar\oint_\gamma \hat{\cal A}_i dx^i   -\hbar^2 (\oint_\gamma \hat{\cal A}_i dx^i)(\oint_\gamma \hat{\cal A}_j dx^j) +\cdots
\ee

In the sequence, we notice that  the  $SU(2)$  gauge potential  is   a  connection  1-form  written in the adjoint representation of the  $SU(2)$   Lie algebra\footnote{
 Using Greek  indices for  space-time, the  corresponding  Yang-Mills curvature     is
a   2-form
$\mathcal{F}= \frac{1}{2} \mathcal{F}_{\mu\nu}\,  dx^\mu \wedge dx^\nu$
where
$\mathcal{F}_{\mu\nu}  =[\mathcal{D}_\mu,\mathcal{D}_\nu ]$, $D_\mu =\partial_\mu  +  \mathcal{A}_\mu$. The potential $ \mathcal{A}_\mu$ is  a  solution of the  Yang-Mills   equations
$\mathcal{D}\wedge \mathcal{F}^{*} =-4\pi J^{*}$  and  $\mathcal{D}\wedge \mathcal{F} =0$,  where $\mathcal{F}_{\mu\nu}^*=\epsilon_{\mu\nu\rho\sigma}\mathcal{F}^{\rho\sigma}$  denotes  the components  of the   dual  $\mathcal{F}^*$.
}
as   ${\mathcal{A}} ={\mathcal{A}}_{\mu}dx^{\mu}$.
On the other hand  this  1-form  can  also  be  written  as  a  quaternion  function  in  the Pauli basis  as  $\mathcal{A} = da_\mu  \sigma^\mu$,  so that   $\mathcal{A}_\mu dx^\mu=da_\mu  \sigma^\mu$, where   $da_\mu$ is  a 1-form  and   $\mathcal{A}_\mu$ are quaternion  functions.  The  3-dimensional curvature
operator (7)  can be  written   explicitly,  using the    3-dimensional components   $\mathcal{A}_i$  of the   imaginary part of a  quaternion function
\be
Im\, {\mathcal{A}}(X)  =\frac{1}{2}(f(X)dX -\overline{f(X)dX})    \label{eq:A}
\ee
Therefore, it is possible  to  to  obtain the   quantum  curvature \rf{R}  by  specifying a  quaternion function expressed in the Pauli  basis  \rf{Pauli}. As  an example, consider   the quaternion  function
\be
f(X)  = \frac{\bar{X}}{1+|X|^{2}}  \label{eq:f}
\ee
and its  quaternion conjugate $\bar{f}(X)=X/(1+|X|^{2})$,
we  may   obtain
self-dual (instanton)  and  anti-self-dual (anti-instanton) $SU(2)$  gauge fields  respectively,   defined by  $\mathcal{F^*}  =\pm \mathcal{F}$.  In fact,  replacing the  above  functions in \rf{A}   we obtain
\be
Im\mathcal{A}(X)  =\frac{1}{2}\left(\frac{\bar{X}dX-\bar{dX} X}{1+  |X|^2}\right)  \label{eq:instantonA}
\ee
and the    gauge   curvature
\be
\mathcal{F}(X) =  \frac{\bar{dX}\wedge  dX}{(1+ |X|^2)^2}  \label{eq:F}
\ee
This  can be readily    verified  to satisfy the   anti-self-dual    (anti-instanton) condition \cite{Atiyah}.
More specifically, writing   $X= x^\mu \sigma_\mu$ and   $dX= dx^\mu \sigma_\mu$  in \rf{instantonA}, then we obtain
\be
\mathcal{A}(X)  =\frac{- x^i dx^0 + x^0 dx^i }{1+  |X|^2}\sigma_i =
 \mathcal{A}_0 dx^0 +  \mathcal{A}_i dx^i \label{eq:AA}
\ee
where
\be
\mathcal{A}_0=-\frac{x^i\sigma_i}{1 + |X|^2}\;\;\mbox{and}\;\; \mathcal{A}_i = \frac{x^0\sigma_i}{1 + |X|^2}  \label{eq:Ai}
\ee
are the components  of the $SU(2)$  connection 1-form.
Replacing the   components   $\mathcal{A}_i$     in  \rf{R} we obtain  the  quantum quaternion exponential  curvature operator  generated  by an  anti-instanton
\be
\hat{\mathcal{R}}({\cal A}) =   \mathcal{P}e^{i\hbar \sigma_i\oint_\gamma \frac{x^0}{1 + |X|^2}dx^i }, \;\; i=1..3  \label{eq:instanton}
\ee
By  taking  the   quaternion conjugate   of \rf{f}  we obtain the  self-dual  (instanton) solution.

\begin{center}
\textbf{Concluding Remarks}
\end{center}
In general  relativity the observables of the gravitational  field  are  associated  with  the  eigenvalues of the curvature  tensor.    Therefore,  the quaternion expression  \rf{R}  may  represent  an observable  effect of the gravity  induced by the quantum  $SU(2)$  gauge  field.
Such quantum  effects  should  be  observed at the same   electro-weak  energy level  of the  generating gauge  potential (at the Tev scale),  so that  in principle   it  can  be
experimentally  verified  at the   LHC in the form  of  small  variations of  the Minkowski   zero   curvature  at the quantum scale.  However,  such result strongly  contrasts    with the
hypothesis  that  the  discrete structure  of the standard loop  quantum gravity  is  defined  at the  Planck  scale.

As it  was  noted, quaternion loop  quantum gravity  differs
fundamentally   from  the  standard loop quantum gravity, because
in the quaternion approach the
existence of a  classical background, the space-time geometry, is  given  by  Einstein's  theory.
In  a sense the  quaternion loop quantum gravity is  a less ambitious program than the  standard  loop  quantum gravity proposition. Here the   classical dynamics of the  gravitational  field  is given by  Einstein  equations,  providing the  foliation of  space-time by  3-dimensional  hypersurfaces.  However,     we also  obtain a   quantum  gravitational component,    given  by  the  Wilson  curvature  \rf{R} which is  effective only at the  quantum scale,  as  derived   from the  time  dependent  $SU(2)$ Yang-Mills gauge  field    equations.

As  a  final remark,   we  may add  that the  quaternion
loop quantum gravity does not depend on the  construction of  a gravitational Hamiltonian or,  indeed  of  a
traditional  canonical  formulation. Therefore it seems that it is possible to  maintain the  diffeomorphism invariance of the classical theory,  even  with the  3+1  decomposition of the  space-time.  This  has been   also  suggested  in the traditional loop  quantum  gravity,  although admittedly  it is  far from obvious \cite{Ashtekarstar}.

\end{document}